\documentstyle[aps,epsf,twocolumn,prl]{revtex}

\begin{document}
\draft

\twocolumn[\hsize\textwidth\columnwidth\hsize\csname @twocolumnfalse\endcsname
\title{The Finite Temperature  Mott Transition  
in the Hubbard Model in Infinite Dimensions}
\author{ Marcelo J. Rozenberg}

\address{Departamento de F\'{\i}sica, FCEN, Universidad de Buenos Aires,
Ciudad Universitaria Pab.I, (1428) Buenos Aires, Argentina.}

\author{ R. Chitra and   Gabriel Kotliar} 
\address
{Serin Physics Laboratory, Rutgers University, Piscataway, NJ 08854, USA.}

\date{\today}
\maketitle
\widetext
\begin{abstract}
\noindent

We study the second order
finite temperature Mott transition  point
in the fully frustrated
Hubbard model at half filling, within Dynamical Mean Field Theory.
Using  quantum Monte Carlo simulations 
we show the existence of a finite temperature
second order critical point by explicitly demonstrating the
existence of   a divergent
susceptibility
as well as by finding coexistence in the low temperature phase.
We determine the   location of the finite temperature Mott 
critical point in the $(U,T)$ plane. 
Our study verifies   and   quantifies   a scenario for the
Mott transition proposed in earlier studies \onlinecite{review} of
this problem.
\end{abstract}


]

\narrowtext

When the ratio of the strength of the electron electron interaction
to the bandwidth is increased, a metal insulator transition (MIT)
occurs\cite{mott}. This phenomenon has been continuously and intensively
studied in narrow bandwidth systems for several decades \cite{ift}, with
$V_2O_3$ being the archetypal system.

Providing a detailed
theoretical description of this transition in
systems
which are not  magnetically  ordered   is one of
the most challenging problems in condensed matter physics.
In recent years, great progress was achieved  using the
 dynamical mean field theory (DMFT), a method which
 becomes exact in the limit of large lattice coordination.
Within this framework,  one can describe  both metallic and insulating
phases.  See \cite{review} for a review.
Though some aspects of this theory
agree with the scenarios put forth earlier by Brinkman
and Rice and by Hubbard \cite{hubbard,br}
the DMFT approach suggested
important   qualitative modifications of these pictures.

At zero temperature, in the DMFT of the
fully frustrated Hubbard model,
 upon increasing the interaction
the paramagnetic metallic state is destroyed
at a critical value denoted by
$U_{c2}$. 
On the other hand,
the  insulating
solution that exists for large $U$
 disappears  when  $U$ is  decreased,
at a different
critical value $U_{c1}$, at which the gap closes.
Since $U_{c1} < U_{c2}$ there is  a
region where  two phases i.e.,  a paramagnetic
metal and a paramagnetic insulator coexist.
This region of coexistence
 naturally extends to finite temperatures, 
where it is bounded by two lines $U_{c1}(T)$  and $U_{c2}(T)$.
Consequently, at finite
temperatures, 
a {\em first order} metal to insulator transition takes
place at a  value of the interaction $U_c(T)$, with
$U_{c1}(T) < U_c(T) < U_{c2}(T)$ \cite{uc2,ucs}.

At  temperatures above this transition line,
two different crossover regions  were identified \cite{thomas}.
The first crossover region, is  the natural  continuation of
the $U_{c2}(T)$ line, and was characterized as the place where the low energy
resonance in the spectral function rapidly loses its intensity and the
resistivity  increases rapidly.
The second crossover region can be considered to be the natural extension
 of the $U_{c1}(T)$, and was characterized as the
line where the gap is comparable to the temperature and where a
crossover to activated behavior in the resistivity is seen.
 These crossover lines  were further related
\cite{thomas} to the experimental
observations of  Mc Whan et al. and Kuwamoto et al. \cite{macwhan,kuwamoto}.
Note that generically within DMFT, i.e.,
both at high and low temperatures ,
the  destruction of the metallicity and the gap closure occur  at different
locations on the phase diagram.
The only  exception to this  is the finite temperature second order critical
point at which  the two boundary lines
 $U_{c1}(T) $ and $U_{c2}(T) $ meet and  terminate.
In this paper,
we provide a  description of this  second order point at which
the MIT occurs. 

Though substantial evidence indicated \cite{review} 
that the qualitative aspects of the finite temperature MIT scenario
that is  described above   were genuine properties
of the exact solution of the model.
Recent  work 
quantum Monte Carlo (QMC) work
challenged
this scenario \cite{superprod}. In Ref.\onlinecite{superprod} 
it was claimed that, within the temperature range numerically accessible
to the QMC technique, 
it was not
possible to find any evidence for a  {\it  finite temperature }
MIT point.
It is very important to clarify this issue, since
our earlier work 
has
successfully predicted new experiments in $V_2 O_3$ \cite{thomas} 
and provided a natural  interpretation of earlier data
\cite{macwhan,kuwamoto}. 
We describe below, how the MIT point can be found and
we determine its location
with reasonable accuracy, illustrating the
power, but at the same time the subtleties of the QMC technique
when applied in conjunction with the self-consistent DMFT
equations.

Besides quelling  the doubts raised recently in Ref.\onlinecite{superprod}
there are other 
motivations for this work. 
As  stressed
  by Nozi\`eres \cite{nozieres},
the Mott transition
lacks an obvious order parameter, so it is not clear {\it
a priori}
which quantities should exhibit singular behavior.
Secondly,
the second order  Mott end point
is observed experimentally in $V_2 O_3$ and $Ni Se_x S_{2-x}$, and
its properties are the subject of ongoing experiments. 
Finally, it is important from the point of view of numerical studies
of correlated electron systems. 
Since a full   analytic solution
to  this problem  is unavailable,
the understanding of models in the limit of large lattice coordination
 requires numerical work and analytic approximations. This is a common 
situation in the field of correlated electron systems and
the Hubbard model in infinite dimensions 
provides an excellent playground
to test  the merit of various
analytic approximations and numerical methods.

We consider the Hubbard model on the Bethe lattice
in the paramagnetic phase \cite{footnote} with
coordination $d$ and hopping $\frac{t}{\sqrt{d}}$ 
in the  $d \to \infty$ limit \cite{metzvoll}
 at half filling.
\begin{equation}
H = - \sum_{ <i,j> \sigma \omega} {\frac{t}{\sqrt{d}}} (c^{+}_{i \sigma}
c_{j \sigma} + c.c.) + \sum_{i} U n_{i \uparrow }n_{i \downarrow}
\end{equation}
The half-bandwidth is given by  $D = 2 t$
and we set $D=1$ as a unit of energy. The chemical potential is 
set equal to $\frac U{2}$ at half-filling.

All the {\em local} correlation functions
can be obtained from a Single Impurity Anderson model (SIAM)
with  a hybridization
function
\begin{equation}
\Delta (i \omega_n) = \sum_{k} \frac {{V_k}^{2}}{i \omega_n-\epsilon_{k}}
\label{delta}
\end{equation}
provided that $\Delta (i \omega_n)$ obeys the self-consistency condition
\cite{ipt}:
\begin{equation}
{t^{2}} {G_{imp}}(i \omega_n) {[\Delta, \alpha]} = \Delta (i \omega_n).
\label{mf}
\end{equation}
Here $\alpha$ denotes the control parameters of the problem  $U$  and  $T$.
$G_{imp} (i \omega_n) [\Delta]$ is the finite temperature 
Matsubara $f$-electron
Green's function of the SIAM
\def\sumks{\sum_{k \sigma }}
\begin{equation}
H_{SIAM}=
\sum_{k \sigma } \epsilon_{k} c^{+}_{k \sigma} c_{k \sigma} + 
V_{k}(c^{+}_{k \sigma}f_{\sigma}  + f_\sigma^{+} c_{k \sigma}) + U 
n_{f{\uparrow}} n_{f{\downarrow}} 
\label{siam}
\end{equation}

To solve the problem, the single impurity Green's function is
calculated using QMC or exact diagonalization methods
and  (\ref{mf}) is used to check for self consistency \cite{review}.
This  iterative process  is continued until self consistency
is achieved. At that point, ${G_{imp}}(i \omega_n)$ coincides
with ${G_{loc}}(i \omega_n)$, the {\em local} Green's function of the
original {\em lattice} problem.
At each iteration step we use
quantum Monte Carlo simulations \cite{fyehirsch} to calculate  the impurity 
Green's function in imaginary time \cite{crowd}.
This algorithm, is widely considered to provide 
numerically exact solutions to the model (in the Monte Carlo sense).

The parameters of the simulation are as follows:
we typically perform 60,000 sweeps (1 sweep = 1 attempt to update
each of the $L$ pseudo-spins = $L$ attempts where $L$ is the number of
time slices used), and  when required upto 300,000 sweeps  
in order
to minimize the enhanced fluctuations close to the critical point.
We always set the length of the time-slice $\Delta \tau = \beta / L \le 0.5$,
where $\beta$ is the inverse temperature.

As a criterion for the convergence of the numerical
solution of the DMFT equations, we monitor the evolution with
the iteration number of $G(i\omega_1)$, the value of
the Green's function at the first Matsubara frequency. This quantity
is appropriate, as it is essentially the integral of $G(\tau)$
(with $\tau$ being the imaginary time), the quantity that
is directly computed statistically in the QMC calculation.  
Also, $G(i\omega_1)$ is the value of the frequency dependent
Green's function that
fluctuates  most, thus, it sets an upper bound for the statistical
error of the whole $G(i\omega_n)$.
We stop the iterations when
the fluctuations in $G(i\omega_1)$ 
become of the order of the statistical error of the QMC
(which is controlled by the number of sweeps at a  given $U$ and $T$) 
and remain stable for at least about 30 more iterations.

In generic regions of the
$(U,T)$ plane less than 10 iterations are sufficient to
obtain a converged solution. 
However, since we are looking for a critical point which is a
bifurcation point of (\ref{mf}),
the number of iterations
needed to obtain convergence diverges,
as we approach the critical point. 
This, is the  usual phenomenon of
critical slowing down associated with bifurcating solutions
in recursive procedures and it forces us to increase the number
of iterations, as we approach the critical point.
For example,  up to 300 iterations were necessary  to check the convergence
at parameter values  close to the critical point.
Special care was taken to assure that the solutions were indeed
converged. To this end, we used two different initial seeds for the
iterative procedure: $G(\tau)[U=0]$ that corresponds to a metallic
state, and a $G(\tau)[t=0]$ that corresponds to an insulating one.
These two initial Green's functions have qualitative different behavior
a low frequencies, one with a finite density of
states at the Fermi energy and the other with an insulating gap.
If there is a single solution to the mean field equations, the
algorithm should in both cases
evolve towards the unique attractor. If two different 
solutions are allowed, one metallic like and one insulator like, it
may be expected that the metallic seed would evolve towards the former
and the insulating seed towards the latter. 
Using this method, when 
we find a converged unique solution to (\ref{mf}), we are
certain that  the algorithm has fully converged.

To prove the existence of a finite temperature critical point
we need to isolate a physical quantity that exhibits singular behavior.
We  view this MIT  as a liquid gas transition, where
the double occupancy $\langle d \rangle$
plays the role of the density, while $U$ plays the
role of the pressure and $T$ the natural role of temperature.
This analogy was put forward in an insightful
paper by Castellani {\it et al.}
\cite{castellani}.  As shown here  their scenario 
is  realized within the DMFT solution of the  Hubbard model.
With this analogy to the liquid gas transition in mind, we 
focus on the behavior of the double occupation as a function of
temperature and interaction strength. We define the susceptibility
$\chi = max_{U} [ \partial \langle d \rangle / \partial U ]$ 
that, as we shall later show,
is a quantity that diverges at the finite $T$ second order critical point.

\begin{figure}
\epsfxsize=3.5in
\epsffile{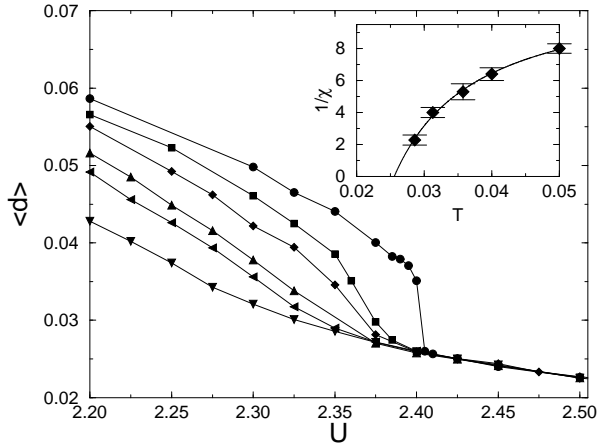}
\epsfxsize=3.5in
\caption{The double occupation $\langle d \rangle$ as a function of $U$ for
several values of $T$= 1/20,  1/25,   1/28  1/32,  1/35,  1/40
(from bottom to top).
The inset shows the inverse susceptibility $\chi^{-1}$ as a function
of $T$. The line is a least squares fit using the expression in the
text. The intercept with the $T-$axis gives our estimate for $T_c=0.026$
}
\label{fig1}
\end{figure}

In Fig. \ref{fig1}, we show our results for the double occupation
$\langle d \rangle$ as a
function of $U$. Each set of data is obtained at a constant $T$, 
and we have checked 
the uniqueness of the solutions using the procedure described above.

The most remarkable feature is the rapid variation of $\langle d \rangle$
with the interaction $U$ at the lower temperatures \cite{ucs}. 
This variation is a direct
consequence of a divergence in the susceptibility $\chi$ 
defined earlier. 
In order to obtain  the susceptibility from the
discrete set of data at each temperature,
we fit the numerical results with analytic expressions that follow from
a Landau  Ginzburg analysis \cite{notes}  to be discussed elsewhere.
We obtain the curve
$\chi^{-1}(T)$ (which is plotted in the inset of Fig.\ref{fig1}), 
that can be approximated, close to the transition,
with the expression $\chi^{-1} = \frac{t}{a+bt}$ 
where $t$ is the reduced temperature $t= T-T_c$ and $a$ and $b$ are real
positive parameters.
More precisely  $\langle d \rangle (U, T) $ 
contains a singular part (arising from the liquid gas analogy close
to the transition) and a regular part.
The regular part gives the constant  background ($b/a$) and the singular part
produces the divergent susceptibility ($a/t$)
which is linear in the inverse reduced temperature. 
$T_c$ and $U_c$ are fitting parameters, which give the
location of the second order critical point where the first
order lines terminate. A least squares fit to $\chi$ gives
our estimate for $T_c = 0.026 \pm 0.003$ and using the liquid gas
analogy we find that $U_c = 2.38 \pm 0.02$. 
Thus, we have obtained
estimates for the  position of the finite
temperature second order critical point
 in the single band
Hubbard model
using the QMC method \cite{goetz} \cite{diag}.
It is important to realize now 
that the lowest temperature set of data for the 
double occupation in Fig.\ref{fig1} 
at $T=1/40=0.025$ appear to be
slightly {\em below} $T_c$. These results
thus indicate a {\em discontinuous} jump in the double occupation.

\begin{figure}
\epsfxsize=3.5in
\epsffile{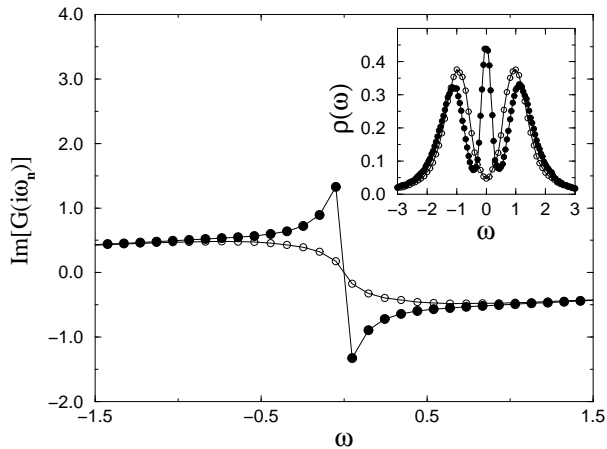}
\epsfxsize=3.5in
\caption{Two converged solutions, one metallic-like (solid circles) and one
insulating-like (open circles), of the DMFT equations for the same value
of the interaction $U=2.4$ and $T=1/64$. In the inset we show their
corresponding density of states. 
}
\label{fig2}
\end{figure}

A more general consequence of the presence of the
finite temperature second order critical point, is the existence of
a  region at $T$ lower than $T_c$,
where two different solutions of the mean field
equations exist. This coexistence of solutions, as usual, results in 
first order transition lines.
We have, therefore, searched for different converged solutions 
well below $T_c =0.026$
and performed calculations at $T=1/51.2$ and $T=1/64$ for several
values of $U$ close to $U_c$. 
Indeed, we obtained two different and
fully converged solutions at $U=2.4$ in the first case, and at $U=2.4$ and
$U=2.5$ in the latter.  For the $U=2.4$ and $T=1/64$ solutions,
we have taken special care to rule out the possibility of systematic
errors, by performing a large number of further iterations 
after self-consistency
was achieved and also  by increasing the number 
of sweeps to up to $300,000$ to
minimize statistical fluctuations.
In Fig.\ref{fig2}, we display two coexistent solutions at $U=2.4$ and 
$T=1/64$, along with the results for
the imaginary part of the Green's functions on the real frequency axis
using the maximal entropy method \cite{mem} for the analytic continuation.
These results, obtained within a narrow range of interaction $U$ close
to $U_c$ and at temperatures {\em below} our estimate
for $T_c$,  further confirm the consistency of our numerical
results.

In conclusion, we presented the results of a careful  numerical 
QMC study of the finite temperature Mott transition in the 
paramagnetic phase of the Hubbard model in infinite dimensions.
We identified  
the singular behavior of a  susceptibility
associated with the  finite temperature critical point
and good  estimates for $U_c$ and $T_c$ were 
obtained. 
This should be contrasted with a previous QMC study of the
same  problem \cite{superprod}
which reached the conclusion 
that there is no signature of a finite
temperature  second order phase transition 
in the temperature region accessible to QMC studies.
Furthermore 
they   presented 
 $T_{bound} = 0.01429 D$  as a  rigorous numerical upper bound on  $T_c$. 
Here we have shown using two different approaches, i.e.,  a high temperature
susceptibility calculation and  an explicit demonstration of coexistence at
low temperatures, that  a finite temperature critical point exists
and that  the  bound  obtained in \cite{superprod} is incorrect.

In earlier publications we asserted that   qualitative
features of the  paramagnetic metal to paramagnetic
insulator transition  observed in  $V_2 O_3$,
are well described by the DMFT of the single band Hubbard model on 
a frustrated lattice treated using the iterated
perturbation theory.  In this work we show that those
conclusions are also valid when a more accurate technique,
such as QMC, is used to solve the dynamical mean field equations.
In particular, the anomalous temperature dependence of physical
properties  (double occupancy, resistivity etc.) and the
crossovers  discussed in Ref. 
\onlinecite{thomas}, which were
tied to the proximity  to a  second order   
finite temperature metal to insulator transition in the model.

While the qualitative agreement between the DMFT and the
experiments is remarkable, there are quantitative disagreements.
The  measured plasma frequency is a factor of two larger than
the calculated one  (using exact diagonalization and DMFT),
if we
take  the  LDA  estimate of the bandwidth  $D \approx 0.5$eV \cite{matheiss}.
This was already noticed in Refs. \cite{thomas,prb}.
Using the same  LDA based  estimates for the model parameters,
the critical temperature that we compute
(with DMFT and QMC) is also approximately a factor of two
smaller than the observed  $T_c$ in $V_2 O_3$ \cite{macwhan}.

These  results suggest the necessity for
the  inclusion of  additional features  like orbital degeneracy,
ligand  bands\cite{wolenski} and the electron phonon interaction\cite{krish}
in the model to be able to make accurate  predictions 
for physical quantities in this system.
The extension of the QMC algorithm to  orbitally degenerate models
large $d$ models 
proposed by one of us \cite{twoband} should be useful for this purpose.
Indeed, less extensive studies than those carried out already show that
orbital degeneracy affects 
the critical temperature \cite{twoband}.

ACKNOWLEDGMENT
Useful conversations with G. Moeller  and W. Krauth are acknowledged.
This work was supported by NSF 95-29138.
MJR acknowledges support of Fundaci\'on Antorchas,
CONICET (PID $N^o4547/96$), and ANPCYT (PMT-PICT1855).

\begin{thebibliography}{99}
\bibitem{footnote}
As discussed in Ref.\onlinecite{review} these results become exact
in many magnetically frustrated lattices 
which share the same density of states as the Bethe lattice in the limit
of infinite coordination.
\end{thebibliography}
\end{document}